\documentclass[prx,preprint,showpacs,superscriptaddress,nofootinbib,amsfonts,amsmath,floatfix]{revtex4}

\usepackage{graphicx}
\usepackage{color}
\usepackage{import}
\usepackage{wrapfig}
\usepackage{mathtools}
\usepackage{bm}
\usepackage{amsmath}
\usepackage{amsfonts}
\usepackage{amssymb}

\begin{document}

%=======================================================================================

\title{Theory of classical electrodynamics with topologically quantized singularities as electric charges}

\author{Bruno Golik}
\altaffiliation{These authors contributed equally to this work.}
\affiliation{Department of Physics, Faculty of Science, University of Zagreb, Bijeni\v{c}ka c. 32, 10000 Zagreb, Croatia}

\author{Dario Juki\'{c}}
%\email{djukic@grad.hr}
\altaffiliation{These authors contributed equally to this work.}
\affiliation{Faculty of Civil Engineering, University of Zagreb, A. Ka\v{c}i\'{c}a  Mio\v{s}i\'{c}a 26, 10000 Zagreb, Croatia}

\author{Hrvoje Buljan}
\email{hbuljan@phy.hr}
\affiliation{Department of Physics, Faculty of Science, University of Zagreb, Bijeni\v{c}ka c. 32, 10000 Zagreb, Croatia}
\affiliation{The MOE Key Laboratory of Weak-Light Nonlinear Photonics, 
TEDA Applied Physics Institute and School of Physics, Nankai University, Tianjin 300457, China}

\date{\today}

\begin{abstract}
We formulate a theory of classical electrodynamics where the only admissible electric charges 
are topological singularities in the electromagnetic field, and charge quantization is accounted 
by the Chern theorem, such that Dirac magnetic monopoles are not needed. 
The theory allows positive and negative charges of equal magnitude, where 
the sign of the charge corresponds to the chirality of the topological singularity. 
Given the trajectory ${\bf w}(t)$ of the singularity, one can calculate electric and magnetic fields 
{\em identical} to those produced by Maxwell's equations for a moving point charge, 
apart from a multiplicative constant factor related to electron charge and vacuum permittivity.
The theory is based on the relativistic Weyl equation in frequency-wavevector space, 
with eigenstates comprising the position, velocity, and acceleration of the singularity, 
and eigenvalues defining the retarded position of the charge. 
From the eigenstates we calculate the Berry connection and the Berry curvatures, 
and identify the curvatures as electric and magnetic fields. 
\end{abstract}

%03.50.De Classical electromagnetism, Maxwell equations
%03.65.Vf Phases: geometric; dynamic or topological

\maketitle

\begin{figure}
\includegraphics[width=1.0\textwidth]{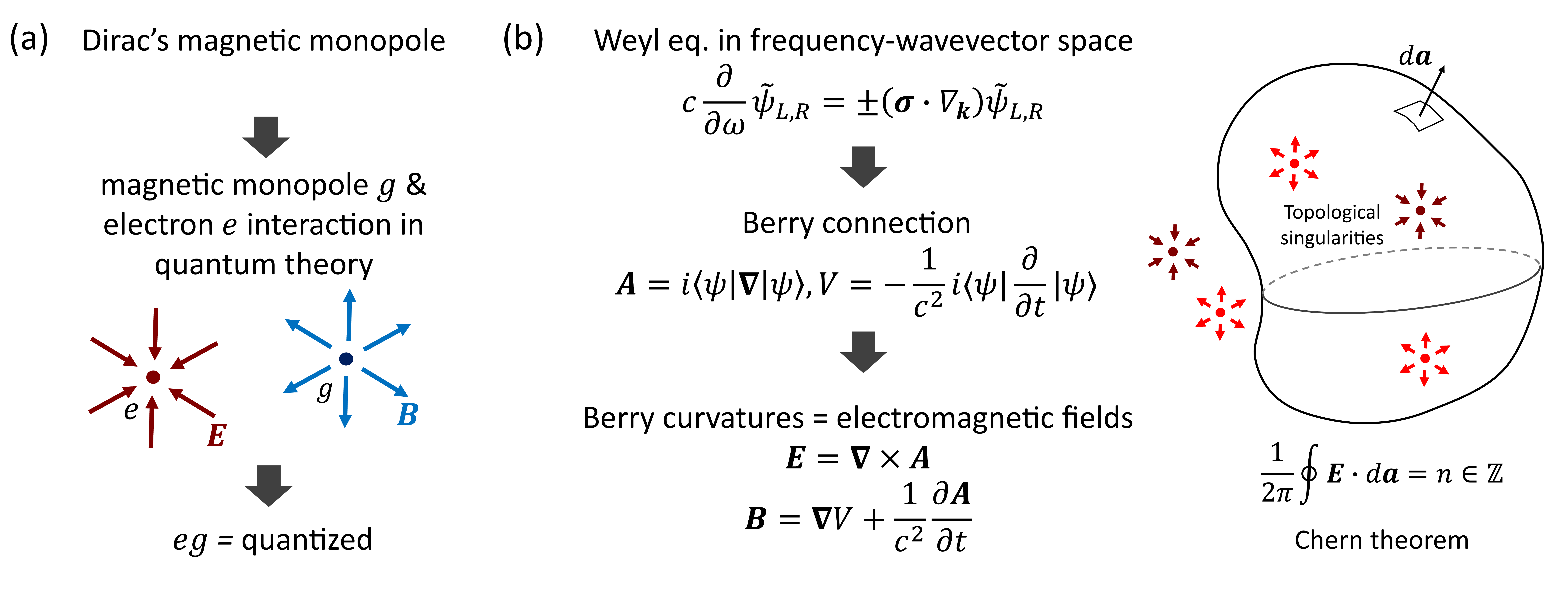}
%\captionsetup[figure]{justification=justified, singlelinecheck=off} 
\caption{Comparison of Dirac's quantization condition and the present theory. 
(a) If a single magnetic charge $g$ (i.e., magnetic monopole) exists in our universe, 
then interaction of the magnetic charge $g$ and an electron $e$ in quantum theory 
leads to quantization of the product $ge$~\cite{Dirac1931,Goldhaber1990}.
(b) In the present theory we construct electric and magnetic fields 
by using Berry connections and curvatures from the solution of 
Weyl equation in frequency-wavevector space. 
These solutions are identical to the solutions of Maxwell's equations for moving point charges, 
obtained via the Li\'{e}nard-Wiechert potentials;
however, Chern theorem plays the role of Gauss law and guarantees quantization of charges, 
which are here topological singularities in space. 
}
\label{fig:1}
\end{figure}

Quantization of charge is one of the long standing unresolved questions of theoretical physics. 
Dirac discovered that if there exists a single magnetic monopole in our universe, this 
would account for the quantization of charge~\cite{Dirac1931}. By exploring behavior of 
an electron in the presence of the magnetic monopole in the realm of quantum mechanics, Dirac  
found that the product of the electron and magnetic monopole charge has to be quantized~\cite{Dirac1931,Dirac1948}. 
This appealing idea attracted great interest including hypothesizing the existence of dyons, 
that is, elementary particles carrying both electric and magnetic charge~\cite{Schwinger1969, Witten1979, Wilczek1982}. 
The existence of magnetic monopole structures was discovered in classical non-Abelian gauge theories 
with spontaneously broken gauge symmetry by 't Hooft~\cite{tHooft1974} and Polyakov~\cite{Polyakov1974}.
However, magnetic monopoles and dyons were not experimentally found 
(for a review of literature on magnetic monopoles see~\cite{Goldhaber1990}).
Here we attempt a more conservative approach and ask the following question: 
Can we develop a theory that can account for the quantization of charge, which would 
simultaneously be consistent with currently accepted theories and experiments?
Towards this goal we formulate a theory of classical electrodynamics where the only admissible electric charges 
are topological singularities in the electromagnetic field. 
There are only two opposite - positive and negative - values of the charge, 
which correspond to the two values of chirality of topological singularities. 
Continuous distributions of charges are not possible, whereas quantization of charge is guaranteed 
by the Chern theorem~\cite{Chern1944} (Fig. \ref{fig:1}). 
We point out that it may be possible to account for charge quantization 
by the {\em classical} theory of electrodynamics. If in some future work this theory is quantized, 
and if the quantum theory will yield identical result as conventional quantum electrodynamics, 
this would account for charge quantization and answer the question posed above. 

Maxwell's theory of classical electrodynamics is conventionally formulated with  
continuous distributions of charges and currents, which are sources of electric and magnetic 
fields~\cite{Griffiths2013, Jackson1999}. During the development of quantum electrodynamics, 
several different formulations of classical electrodynamics emerged including 
Lorentz-Dirac~\cite{Dirac1938} and Wheeler-Feynman formulation
\cite{Wheeler1945, Wheeler1949}, e.g., see Ref.~\cite{Beil1975} and refs. therein. 
A nice account of the problems faced, especially that of infinities related to self-energy 
of a point charge, and the line of reasoning used in developing the theory, 
can be found in Feynman's Nobel lecture~\cite{Feynman1966}. 
However, these theories were not formulated to address the problem of charge quantization.

According to Maxwell's theory, the scalar and the vector potential 
for a point charge $q$ moving along a trajectory ${\bf w}(t)$ 
are the Li\'{e}nard-Wiechert potentials~\cite{Griffiths2013, Jackson1999}:
\begin{equation}
V_{M}(x^\mu)=\frac{1}{4\pi\epsilon_0}\frac{q c}{s c-{\bf s}\cdot {\bf v}},\ \
{\bf A}_{M}(x^\mu)=\frac{{\bf v}}{c^2}V_{M}.
\label{LW}
\end{equation}
Here, $x^\mu=(ct,{\bf r})$ denotes a point in spacetime, $c$ is the speed of light, 
${\bf s}={\bf r}-{\bf w}(t_r)$, $s=|{\bf s}|$, and 
${\bf v}=\frac{ d {\bf w}(t)}{dt}|_{t_r}$ is the velocity of the point charge at the 
retarded time $t_r$. The retarded time is implicitly defined by the equation
\begin{equation}
|{\bf r}-{\bf w}(t_r)|=c(t-t_r),
\label{tr}
\end{equation}
which arises from the fact that information on the position, velocity, and acceleration of the 
point charge travels at the speed of light.  
The electric and magnetic fields are given by 
${\bf E}_M(x^\mu)=-\nabla V_M-d {\bf A}_M/dt$, and 
${\bf B}_M(x^\mu)=\nabla \times {\bf A}_M$.
The lower index $M$ stands for Maxwell's equations.

The formulation of classical electrodynamics presented here yields identical 
expressions for the electric and magnetic fields of a point charge (except for a multiplicative constant), 
however, the scalar and the vector potentials considerably differ. 
A fundamental difference between Maxwell's equations and our theory is the following:
Maxwell's equations allow solutions with continuous distributions of charges and currents, 
while the present theory allows only point charges with two opposite 
(positive and negative) values of the charge.

The theory is based on the relativistic Weyl equation \cite{Weyl1929} in 
frequency-wavevector (that is, energy-momentum) space:
\begin{equation}
c\frac{\partial}{\partial \omega}\tilde\psi_R(k^\mu)
=-
(\sigma_x\frac{\partial}{\partial k_x}+\sigma_y\frac{\partial}{\partial k_y}+\sigma_z\frac{\partial}{\partial k_z})
\tilde\psi_R(k^\mu),
\label{Weyleq}
\end{equation}
where $\sigma_i$, $i,j=1,2,3$, are the Pauli matrices, 
and $k^\mu=(\omega/c,{\bf k})$ is a 4-vector in the frequency-wavevector space. 
Equation (\ref{Weyleq}) is the right-handed form of the Weyl equation, hence the index $R$ in $\tilde\psi_R(k^\mu)$. 
The left-handed form is obtained by placing a minus sign on the left-hand side of Eq. (\ref{Weyleq}).
Wavefunction $\tilde\psi_R(k^\mu)$ is a two-component spinor. 
In this theory, we calculate the electromagnetic fields from eigenstates of the 
Weyl equation by using Berry connection and Berry curvature machinery~\cite{Berry1984}, 
where ''parameters'' are space and time. 
Thus, $\tilde\psi_R(k^\mu)$ should be regarded as an auxiliary mathematical field used 
for generating the electromagnetic field. The auxiliary field contains information 
on the trajectory of a moving point charge:
\begin{equation}
\tilde\psi_R(k^\mu)=\psi_R(x^\mu) \exp(i {\bf k}\cdot {\pmb \rho} - i \frac{\omega}{c} \rho^0 ),
\label{state}
\end{equation}
where $\rho^\mu=(\rho^0,{\pmb \rho})$ is a displacement 4-vector 
\begin{equation}
\rho^\mu=\Lambda^\mu_\nu (x^\nu-w^\nu).
\label{rho}
\end{equation}
Here, $w^\nu=(ct_c,{\bf w}(t_c))$ denotes the position of the charge at time $t_c$, 
that is, $w^\nu$ describes world-line of the moving charge, 
whereas $x^\nu=(ct,{\bf r})$ is a point in spacetime where we want to know 
the electric and magnetic fields. We did not specify how $w^\nu$ and $x^\nu$ are related, 
as this connection naturally arise from the theory. 
Tensor $\Lambda^\mu_\nu$ in Eq. (\ref{rho}) is a Lorentz transformation that depends 
on the velocity and acceleration of the moving charge in a manner specified below. 

%%%%%%%%% Figure %%%%%%%%%%%%%%%%
%%%%%%%%% Figure %%%%%%%%%%%%%%%%

From Eqs. (\ref{Weyleq}) and (\ref{state}) we obtain an eigenvalue equation 
\begin{equation}
H\psi_R(x^\mu)={\pmb \sigma}\cdot {\pmb \rho}\psi_R(x^\mu)
=\rho^0 \psi_R(x^\mu),
\label{EigenEq}
\end{equation}
which has two eigenstates $|\psi_{R,n} \rangle$ and $|\psi_{R,p} \rangle$, 
with opposite eigenvalues equal in magnitude: $\rho^0=\pm |{\pmb \rho}|$
(this is equivalent to $\rho^\mu\rho_\mu=0$ in 4-vector notation). 
The eigenstate $|\psi_{R,p} \rangle$ corresponds to the positive eigenvalue $\rho^0>0$, 
whereas the negative eigenvalue corresponds to $|\psi_{R,n} \rangle$. 
Because $\rho^\mu$ is obtained from the displacement 4-vector 
$s^\nu=x^\nu-w^\nu$ with a Lorentz transformation, $\rho^\mu\rho_\mu=0$
implies that $s^\mu s_\mu=0$.
Thus, the two events, $x^\nu$ and $w^\nu$, are connected by a signal traveling at the speed of light. 
The Lorentz transformation cannot reverse the time-ordering of two events separated by a light-like interval, 
therefore, $\rho^0>0$ implies $s^0>0$ and vice versa. 

As we have already stated, $w^\nu=(ct_c,{\bf w}(t_c))$ is the world-line of the moving charge, whereas 
we aim to calculate the electromagnetic fields at $x^\nu=(ct,{\bf r})$. 
If we rewrite $s^\mu s_\mu=0$ as $c(t-t_c)=\pm |{\bf r}-{\bf w}(t_c)|$, 
we see that the positive (negative) sign in this equation corresponds to 
$t>t_c$ ($t<t_c$, respectively). 
This means that when we use $|\psi_{R,p} \rangle$ together with 
$c(t-t_c)=|{\bf r}-{\bf w}(t_c)|$ to calculate the electromagnetic fields, 
we obtain retarded field solutions. 
In contrast, from $|\psi_{R,n} \rangle$ and $c(t-t_c)=-|{\bf r}-{\bf w}(t_c)|$ 
we obtain advanced field solutions. 
By applying the same procedure for the left-handed form of the Weyl 
equation, we find that its eigenstates $|\psi_{L,p} \rangle$ and $|\psi_{L,n} \rangle$ yield 
the retarded and advanced fields, respectively. 
The positive eigenvalue implies $c(t-t_c)=|{\bf r}-{\bf w}(t_c)|$, which is fully equivalent 
to Eq. (\ref{tr}) with $t_r=t_c$, i.e., positive eigenvalues define the retarded time $t_r$. 
Here we consider only the retarded solutions as they are consistent with causality. 

The Lorentz transformation $\Lambda$ is of the form $\Lambda=RB$, where $B({\bf v})$ is a pure boost by 
velocity of a moving charge ${\bf v}(t_r)$ at the retarded time $t_r$, 
whereas $R({\pmb \theta})$ is a rotation $R({\pmb \theta})=e^{i(\theta_x J_x+\theta_y J_y+\theta_z J_z)}$,
which depends on the velocity and acceleration of the point charge at the retarded time $t_r$: 
${\pmb \theta}(t_r)={\bf \hat n} \cdot \int^{t_r} ({\pmb \omega}_{Th}(t')\cdot {\bf \hat n}) dt'$, 
where 
\begin{equation}
{\pmb \omega}_{Th}(t')=\frac{1}{c^2}\frac{\gamma^2}{\gamma+1}({\bf a}(t')\times {\bf v}(t'))
\end{equation}
is the Thomas precession frequency~\cite{Thomas1927}, and ${\bf \hat n}$ is a fixed 
unit vector aligned with ${\bf a}\times {\bf v}$ at the retarded time $t_r$. 
Here, $J_i$ are generators of rotations, and $\gamma=1/\sqrt{1-v^2/c^2}$ is the Lorentz contraction. 
Later in the text we will demonstrate that the electromagnetic fields depend only on the 
derivatives of the angle variables $d{\pmb \theta}(t_r)/dt_r={\pmb \omega_{Th}(t_r)}$ at $t_r$. 
For a one-dimensional (1D) motion of the charge, ${\bf a}\times {\bf v}=0$ and $\Lambda$ is a pure boost $\Lambda=B({\bf v})$. 
For curvilinear motion, the boost is followed by rotation. 
If the curvilinear motion is confined to a plane, the definition of the angle simplifies to 
${\pmb \theta}(t_r)=\int^{t_r} {\pmb \omega}_{Th}(t') dt'$.

The vector and the scalar potentials in present theory are given by the 
Berry connection~\cite{Berry1984}:
\begin{equation}
{\bf A}=i\langle \psi |\nabla| \psi \rangle,\ \ V=-\frac{1}{c^2}i\langle \psi |\frac{\partial}{\partial t}| \psi \rangle.
\label{AVpot}
\end{equation}

The electric and magnetic fields are given by the Berry curvature:
\begin{equation}
{\bf E}=\nabla \times {\bf A},\ \ {\bf B}=\nabla V+ \frac{1}{c^2}\frac{\partial {\bf A}}{\partial t}.
\label{EBfield}
\end{equation}
In Eqs. (\ref{AVpot}), $| \psi \rangle$ is either $| \psi_{R,p} \rangle$ or $| \psi_{L,p} \rangle$. 
The two eigenstates $| \psi_{R,p} \rangle$ and $| \psi_{L,p} \rangle$ have a well-defined chirality. 
By calculating the direction of the electrostatic field (see below), 
we find that positive charge corresponds to the fields obtained from the eigenvector 
$| \psi_{L,p} \rangle$, whereas fields obtained from $| \psi_{R,p} \rangle$ correspond to the 
negative charge. For this reason, in the rest of the text we use the notation 
$|\psi_{+} \rangle=|\psi_{L,p} \rangle$ and  $|\psi_{-} \rangle = |\psi_{R,p} \rangle$.

Two technical notes are in order:
While evaluating the derivatives in Eqs. (\ref{AVpot}) and (\ref{EBfield}), one should 
take into account the fact that the retarded time $t_r$, which is implicitly defined 
with Eq. (\ref{tr}), depends on the coordinates $x^\nu=(ct,{\bf r})$. 
Therefore, derivatives of ${\bf v}(t_r)$ and ${\bf a}(t_r)$ with respect to spatial 
coordinates $x$, $y$, and $z$ are not zero.
The connection ${\bf A}$ has a singularity fully equivalent to the Dirac string 
of the Dirac magnetic monopole~\cite{Dirac1931}. This was clarified by Wu and Yang, who have shown 
that two connections are required to cover the entirety of parameter space (which is 
real space here); the two connections are related by a gauge transformation 
in a region where they overlap~\cite{Wu1975}.

Although the theory is written for a single charge, by postulating the superposition principle 
it is straightforward to expand it for a number of charges; 
for simplicity we will discuss a single charge. 
Note that in our theory the electric field is a curl of the vector potential ${\bf A}$. 
This means that the only admissible charged objects are singularities of the electromagnetic field 
in space where $\nabla\cdot (\nabla \times {\bf A})=0$ is not applicable. 
Moreover, by construction of the electric field, these singularities are topological, 
that is, Chern theorem guarantees that 
\begin{equation}
\oint_S {\bf E}\cdot d{\bf a}=2\pi \times \text{integer},
\end{equation}
where the integer corresponds to the number and chiralities of the topological singularities
inside the closed surface $S$, see Fig. \ref{fig:1}. 
In our theory, the sign of the charge corresponds to the chirality of the topological singularity
in space.

The electric and magnetic fields in Eq. (\ref{EBfield}) differ from the 
conventional Maxwell fields by a multiplicative constant 
\begin{equation}
{\bf E}_M=\frac{q}{2\pi \epsilon_0}{\bf E}, \ {\bf B}_M=\frac{q}{2\pi \epsilon_0}{\bf B},
\label{multip}
\end{equation}
where $q$ is the electron charge.

First, we demonstrate that Eq. (\ref{multip}) indeed holds. We start with the simplest 
example of a stationary point charge at a position ${\bf w}$, ${\bf v}=0$, where 
Hamiltonian (\ref{EigenEq}) takes the form $H={\bf \sigma}\cdot ({\bf r}-{\bf w})$. 
This is a well-known Hamiltonian that yields a Berry monopole at ${\bf w}$, that is, 
\begin{equation}
{\bf E}=\pm\frac{1}{2}\frac{{\bf r}-{\bf w}}{|{\bf r}-{\bf w}|^3}, \ {\bf B}=0;
\label{EBstationary}
\end{equation}
positive (negative) sign corresponds to fields obtained with 
$| \psi_{+} \rangle $ ($| \psi_{-} \rangle $, respectively).

Next, we consider a point charge moving with the constant velocity along the $z$-axis, 
${\bf w}=v_z t \pmb{\hat z}$, where Hamiltonian (\ref{EigenEq})  takes the form
\begin{equation}
H = x\sigma_x + y\sigma_y + \gamma (z-v_z t)\sigma_z .
\label{Hconstantv}
\end{equation}
By calculating the Berry connections and curvatures (Appendix \ref{constantv_appendix}), we obtain:
\begin{equation}
{\bf E}=\pm\frac{1}{2}\frac{\gamma(\pmb{r}-v_z t\pmb{\hat{z}})}{\left[x^2+y^2+\gamma^2(z-v_z t)^2\right]^{3/2}}, \ {\bf B}=\frac{1}{c^2}v_z \pmb{\hat{z}} \times {\bf E}
\end{equation}
which are exactly the electric and magnetic fields of the point charge moving at a constant velocity. 
This result is perhaps not surprising because we have essentially Lorentz boosted the stationary Hamiltonian 
${\bf \sigma}\cdot ({\bf r}-{\bf w})$ to obtain Eq. (\ref{Hconstantv}).

However, what we find surprising is that this approach yields correct Maxwell expressions 
for the field of a charge that accelerates and radiates. 
First, we address motion of a charge along a straight line, where velocity and acceleration are collinear
(as in Bremsstrahlung).  
Because ${\bf a}\times {\bf v}=0$, we have $R=1_{4}$, and Lorentz transformation in Eq. (\ref{rho}) is 
a pure boost, $\Lambda=B(\pmb{v} (t_r))$. 
Without losing generality, we set the motion of the charge along the 
$z$-axis. The Hamiltonian  (\ref{EigenEq}) is given by 
\begin{equation}
H = x\sigma_x + y\sigma_y + \gamma(t_r) (z-w_z(t_r)-v_z(t_r)(t-t_r))\sigma_z.
\label{H1D}
\end{equation}
Although it is essentially the same Hamiltonian as in the previous example, 
the derivative $dv_z(t)/dt|_{t_r}$ is now not necessarily zero, which gives rise 
to the radiation field. 
By applying the machinery of connections and curvatures (Appendix \ref{1D_motion_appendix}), we find 
the electric and magnetic fields to be:
\begin{align}
{\bf E} & = \pm \frac{1}{2}\frac{s}{(\pmb{s}\cdot\pmb{u})^3}\left[(c^2-v_z^2) \pmb{u} + \pmb{s}\times(c{\pmb{\hat{s}}}\times(a_z{\pmb{\hat{z}}}))\right], \nonumber \\
{\bf B} & = \frac{1}{c}\pmb{\hat{s}}\times {\bf E},
\label{1D}
\end{align}
coinciding with Maxwell's theory. 
Here, we have used $\pmb{u} \equiv c \pmb{\hat{s}} - \pmb{v}$ following~\cite{Griffiths2013}.

Now we turn to curvilinear motion. For simplicity, let us first address motion of a charge in a 
plane, which we choose to be the $xy$ plane without loss of generality. 
In this case, the rotation component of the Lorentz transformation $\Lambda=RB$ 
is of the form $R(\theta_z(t_r))=e^{i \theta_z(t_r) J_z }$, where 
$\theta_z(t_r)=\int^{t_r} \omega_{Th}(t') dt'$,
whereas the boost is $B(v_x(t_r) {\bf \hat x}+v_y(t_r) {\bf \hat y})$. 
The Thomas precession frequency is at any point of motion orthogonal to the $xy$ plane: 
${\pmb \omega}_{Th}(t') = \omega_{Th}(t') {\bf \hat z}$. 
By using $\Lambda=RB$, we construct the Hamiltonian (\ref{EigenEq}) and apply 
Eqs. (\ref{AVpot})-(\ref{EBfield}) to calculate the fields. 
A detailed expression for $\Lambda=RB$ and calculation of the electric field component 
$E_z$ is written in Appendix \ref{2D_appendix}, 
while the other field components can be obtained in an analogous manner
yielding
\begin{equation}
{\bf E} = \pm\frac{1}{2}\frac{s}{(\pmb{s}\cdot\pmb{u})^3}\left[(c^2-v^2) \pmb{u} + 
\pmb{s}\times({\pmb{u}}\times \pmb{a})\right], \
{\bf B} = \frac{1}{c}\pmb{\hat{s}}\times {\bf E}.
\label{2D}
\end{equation}
The resulting fields in Eq. (\ref{2D}) coincide with those given by Maxwell's theory~\cite{Griffiths2013}. 

Next, consider a curvilinear motion in three-dimensions (3D). First, we note that
Eq. (\ref{2D}) coincides with solutions of Maxwell's theory for a 3D curvilinear motion of the charge~\cite{Griffiths2013}. 
The reason behind this is that the electromagnetic fields at a point $x^\mu$ depend on 
the instantaneous velocity and acceleration at the retarded position of the charge. 
The retarded velocity and acceleration vectors are in a plane that is perpendicular to 
the unit vector ${\bf \hat n}={\bf a}(t_r)\times {\bf v}(t_r)/|{\bf a}(t_r)\times {\bf v}(t_r)|$. 
Thus, we can always invent a 2D motion in the plane perpendicular to ${\bf \hat n}$ that has 
identical instantaneous velocity and acceleration at the retarded position of the charge
as a given 3D curvilinear motion, and the electromagnetic fields at $x^\mu$ must be, 
according to Maxwell, identical for the two different motions.
In other words, if we have two world-lines, one describing 2D and the other 
3D curvilinear motion, such that at the retarded position $w^\nu=(ct_r,{\bf w}(t_r))$ 
the velocity and acceleration of the two motions coincide, these two motions will yield 
identical electromagnetic fields at points $x^\mu=(ct,{\bf r})$, 
which are connected with $w^\nu$ by a signal traveling at the speed of light:
$s^\mu s_\mu=0$, $s^\mu=x^\mu-w^\mu$. 
Therefore, as we wish our theory to yield identical solutions as Maxwell's theory, 
we define the unit vector ${\bf \hat n}$ to be fixed perpendicular 
to the plane spanned by the retarded velocity and acceleration, and define the 
angle of rotation as ${\pmb \theta}(t_r)={\bf \hat n} \cdot \int^{t_c} ({\pmb \omega}_{Th}(t')\cdot {\bf \hat n}) dt'$. 
With this definition, we have economically constructed the Lorentz transformation $\Lambda=R({\pmb \theta}(t_r))B({\bf v}(t_r))$ 
that yields correct expressions for the electromagnetic field of a moving charge. 
Beside the general analytical construction described above, we have verified this result numerically as well. 

Let us address the Lorentz covariance of the theory. Suppose that we observe motion of a charge in an 
inertial frame $S$. We insert the world-line of that charge $w^\mu$, which contains the velocity ${\bf v}$ 
and acceleration ${\bf a}$ of the charge in frame $S$ into Eqs. (\ref{Weyleq})-(\ref{EBfield})
that constitute our theory, and obtain the fields ${\bf E}$ and ${\bf B}$ at $x^\mu$ in frame $S$. 
If we move to another inertial frame $S'$, where the world-line $w'^\mu$, 
the velocity ${\bf v'}$, and acceleration ${\bf a'}$ can be obtained via Lorentz transformations, 
and insert these quantities in the same Eqs. (\ref{Weyleq})-(\ref{EBfield}), but now with primed quantities, 
we obtain the fields ${\bf E'}$ and ${\bf B'}$ at $x'^\mu$ in frame $S'$.
We know that the fields ${\bf E}$ and ${\bf B}$ transform into ${\bf E'}$ and ${\bf B'}$ as 
a 2nd rank tensor under Lorentz transformations, simply because they are solutions of 
Maxwell's equations. Therefore, the Lorentz covariance of our theory is connected to the fact that 
we reproduce Maxwell's theory (for point charges). The Weyl equation (\ref{Weyleq}) is manifestly covariant. 

The idea for formulating electrodynamics in terms of the Weyl equation in frequency-momentum space 
arose from studies of the Weyl semimetals in condensed-matter physics, photonics, and 
ultracold quantum gases (e.g., see Refs.~\cite{Turner2013, Lu2015, Xu2015, Lv2015, Dubcek2015}). 
Under specific circumstances, Weyl points may occur in momentum space of 
crystalline, photonic, or optical lattices. 
These momentum space topological singularities are located somewhere in the Brillouin zone(s) of these 
materials. The equivalent of Gauss law for these Weyl points in momentum space is the 
Chern theorem~\cite{Turner2013, Lu2015, Xu2015, Lv2015, Dubcek2015}. 
The idea for this study was simply to exchange the real and momentum space 
to obtain quantized topological charges as electric charges in real space, such that the Gauss law 
in real space would be equivalent to the Chern theorem.

It has been previously shown that Maxwell's equations can be derived from the 
massless Dirac equation in spinor form (e.g., see Refs. \cite{Morgan1995, Rozzi2009} and refs. therein). 
These approaches are equivalent to Maxwell's electrodynamics that allows for 
continuous distributions of charges, and therefore they considerably 
differ from our theory. 
The Berry phase effects have been addressed in momentum space of the 
massive~\cite{Bialynicki1987}, and more recently the massless 
Dirac (i.e., Weyl) equation~\cite{VanMechelen2019}, with Dirac monopoles in 
momentum space. Ref.~\cite{VanMechelen2019} also addressed the monopole 
in momentum space of Maxwell equations. These calculations are in sharp contrast to 
this theory, which addresses electric monopoles in real space 
arising from the Weyl equation in frequency-wavevector space, 
and establishes connection to the fields obtained from 
Li\'{e}nard-Wiechert potentials~\cite{Griffiths2013, Jackson1999}.

Up to this point we have considered the field $\tilde\psi_{R,L}(k^\mu)$ simply as an 
auxiliary field that enforces quantization of charge via Chern theorem, which 
very conveniently yields, via Berry connection-curvature machinery, electromagnetic fields 
of moving point charges. 
One may ask, are these Weyl fields more than a mathematical convenience? 
Can they be interpreted as particles that interact with electric charges and how?
The hypothetical particles described by fields $\tilde\psi_{R,L}(k^\mu)$ are solutions of the 
Weyl equation in frequency-wavevector space, that is, in energy-momentum space. 
Therefore, they live in a space dual to Minkowski spacetime. 
However, dynamics in the two spaces are not independent. 
When a particle moves through spacetime, under the influence of interactions, 
its energy and momentum can change. 
Inversely, if a hypothetical particle described by the field $\tilde\psi_{R,L}(k^\mu)$ 
moves through energy-momentum space, its spatial and temporal coordinates can change. 
For dynamics in such a space, one should find the laws of "conservation of space'' and 
"conservation of time'', which are analogous (or perhaps dual) to conventional 
laws of conservation of momentum and energy, respectively. 
Equation (\ref{AVpot}) would represent interactions of these 
hypothetical $\tilde\psi_{R,L}(k^\mu)$ particles with electric charges. 
Thus, when such a particle moves through spacetime, its wavefunction $\tilde\psi_{R,L}(k^\mu)$ 
acquires a phase fully analogous to the (geometric) Berry phase~\cite{Berry1984}; 
in this case spacetime coordinates can be thought of as parameters, such that a change of 
parameters imprints the geometric phase on the particle's wavefunction. 
This discussion implies that to think of $\tilde\psi_{R,L}(k^\mu)$ as a physical field, 
we should change the standard paradigms of physics such as conservation of energy and momentum. 
Therefore, in this paper we regard $\tilde\psi_{R,L}(k^\mu)$ as auxiliary mathematical fields, 
and leave potential manifestation of their physical reality for future studies. 

In this theory we establish the electromagnetic fields from the world-lines of 
topological singularities (sources), however, the theory does not include the force 
on the singularities, that is, the Lorentz force on point charges. 
The coupling coefficient $q$ is not fixed by the theory, which is related to the 
fact that Maxwell's fields and ours are connected by a constant, see (\ref{multip}). 
However, our theory allows only two opposite values of the charge, 
and they are necessarily point charges, which are constraints on the fields 
not contained in Maxwell's theory. 

In conclusion, we have formulated a theory of classical electrodynamics where 
electric charges are topological singularities in the electromagnetic field, and 
their sign corresponds to the chirality of the singularity. 
Charge quantization is thus accounted by the Chern theorem.
The electric and magnetic fields are {\em identical} to those produced by the 
Maxwell's equations, apart from the multiplicative constant $q/2\pi\epsilon_0$.
The theory is based on the relativistic Weyl equation in frequency-wavevector space, 
with eigenstates depending on the spacetime coordinates. 
From the eigenstates, we calculate the Berry connection and curvatures, by using spacetime 
coordinates as "parameters'', and then identify the curvatures as electric and magnetic fields. 
In outlook, we foresee efforts to develop the quantum version of this theory to 
test whether it will yield identical result as conventional quantum electrodynamics.

We are indebted to I. Smoli\'{c}, M. Solja\v{c}i\'{c}, V. Paar, H. \v{S}tefan\v{c}i\'{c}, Z. Chen, K. Kumeri\v{c}ki, R. Pezer, K. Lelas and I. Kaminer for useful discussions and comments. 
This research is supported by the QuantiXLie Center of Excellence, a project co-financed by the Croatian Government and European Union through the European Regional Development Fund - the Competitiveness and Cohesion Operational Programme (Grant KK.01.1.1.01.0004).

%==============================================================================

\appendix
\section{Charge moving at constant velocity}
\label{constantv_appendix}

In order to unite the calculation of ${\bf E}$ and ${\bf B}$ into one procedure, 
it is useful to define a connection vector $A_\mu$ and an antisymmetric curvature tensor 
$\Omega_{\mu\nu}$ component-wise in the following way:
\begin{equation}
A_\mu = i\langle\psi|\partial_\mu|\psi\rangle,
\label{con_vec}
\end{equation}
\begin{equation}
\Omega_{\mu\nu} = \partial_\mu A_\nu - \partial_\nu A_\mu.
\label{cur_ten}
\end{equation}
Indices $\mu,\nu$ denote components of a chosen coordinate basis. 
We introduce two noteworthy basis choices: $b' = (-c^2t, \rho^1, \rho^2, \rho^3)$ and $b = (-c^2t, x, y, z)$. 
Tensor components evaluated in these bases are denoted using primed and unprimed indices respectively. 
It can be verified that the components of the curvature tensor in basis $b$ correspond exactly to the 
fields given by (\ref{EBfield}) as follows,
\begin{equation}
B_i=\Omega_{i0},\quad
\epsilon_{ijk}E_k=\Omega_{ij},
\label{EB_Omega}
\end{equation}
where $\epsilon_{ijk}$ is the Levi-Civita symbol and Latin indices run from 1 to 3. 
On the other hand, $b'$ contains the natural coordinates of the Hamiltonian  (\ref{EigenEq}): $\rho^1$, $\rho^2$, and $\rho^3$. 
The expressions for the components of connection and curvature will therefore have a very simple form in this basis.

The calculation procedure we use involves first obtaining the connection and curvature in the primed basis, 
after which the unprimed components are calculated using standard basis change tensor transformations. 
Eigenvectors $|\psi_+\rangle$ and $|\psi_-\rangle$ obtained from (\ref{EigenEq}) (and the corresponding left-hand equation) 
are given by:
\begin{equation}
|\psi_+\rangle=
\begin{pmatrix}
\sin{\frac{\theta}{2}}e^{-i\phi} \\ -\cos{\frac{\theta}{2}}
\end{pmatrix},
\;\;|\psi_-\rangle=
\begin{pmatrix}
\cos{\frac{\theta}{2}}e^{-i\phi} \\ \sin{\frac{\theta}{2}}
\end{pmatrix}.
\label{Eigenvectors}
\end{equation}
Here, we used polar coordinates to express $\pmb{\rho}$ as $\rho^1 = \rho \sin{\theta}\cos{\phi}$, 
$\rho^2=\rho \sin{\theta} \sin{\phi}$, $\rho^3=\rho \cos{\theta}$. 
From this point forward, we will consider only results obtained using $|\psi_+\rangle$, as a completely 
analogous calculation can be performed for $|\psi_-\rangle$, with the only difference in the resulting fields being a sign change. 
After inserting $|\psi_+\rangle$ into (\ref{con_vec}) and reverting back to Cartesian coordinates, we obtain the connection components:
\begin{align}
A_{0'} &= A_{3'} = 0, \nonumber\\
A_{1'} &= \frac{1}{2}\frac{-\rho^2}{(\rho^1)^2+(\rho^2)^2} \left(1-\frac{\rho^3}{|{\pmb \rho}|}\right), \label{old_connection} \\
A_{2'} &= \frac{1}{2}\frac{\rho^1}{(\rho^1)^2+(\rho^2)^2} \left(1-\frac{\rho^3}{|{\pmb \rho}|}\right). \nonumber
\end{align}
It should be noted that the acquired connection diverges on the negative half of the $\rho^3$ axis. 
This is a well known result which states that two connections are required in order to cover the entirety 
of the parameter space, e.g., see Ref.~\cite{Wu1975}. 
The second connection is obtained in a different gauge, which is equivalent to multiplying $|\psi_+\rangle$ by a phase factor;
one can choose the gauge for the second connection such that it diverges on the positive half of the $\rho^3$ axis. 
The two connections are then connected by a gauge transformation in a region in space
where they overlap~\cite{Wu1975}.

We now demonstrate the solution for the simplest type of trajectory: 1D motion with constant velocity along the $z$ axis, described by Hamiltonian (\ref{Hconstantv}). Using (\ref{old_connection}) and the relations $\rho^1=x$, $\rho^2=y$, $\rho^3=\gamma(z-vt)$, we obtain the connection in the unprimed basis:
\begin{equation}
A_\mu=\frac{\partial b^{\nu'}}{\partial b^{\mu}}A_{\nu'} \rightarrow
A_{0} = A_{3} = 0, \;
A_{1} = A_{1'}, \;
A_{2} = A_{2'}, 
\label{connection_new_constantv}
\end{equation}
where $b^{\nu'}$ and $b^{\mu}$ denote primed and unprimed basis coordinates respectively. 
Following the curvature definition (\ref{cur_ten}) and identifying the fields via (\ref{EB_Omega}), we obtain the fields:
\begin{align}
\bf{E}&=\frac{1}{2}\frac{\gamma(\pmb{r}-\pmb{v}t)}{\left[x^2+y^2+\gamma^2(z-vt)^2\right]^{3/2}}, \label{E_constv}\\
\bf{B}&=\frac{1}{2c^2}\frac{\gamma v_z\pmb{\hat{z}}\times\pmb{r}}{\left[x^2+y^2+\gamma^2(z-vt)^2\right]^{3/2}}=\frac{1}{c^2}\pmb{v}\times\bf{E}, \label{B_constv}
\end{align}
which are the same as the fields of a constant velocity charge given by Maxwell's equations, 
up to a multiplicative constant $q/2\pi\epsilon_0$.

\section{Charge motion along a straight line}
\label{1D_motion_appendix}

In this section, we present a generalization of the result obtained in Appendix \ref{constantv_appendix} 
by allowing the velocity to vary with time, while still restricting the particle trajectory to a straight line ($z$ axis). 
This Hamiltonian describing such 1D motion is given in (\ref{H1D}).

For the sake of simplicity, we will use a slightly modified version of the calculation method presented in appendix \ref{constantv_appendix}. Instead of transforming the connection vector and using it to calculate $\Omega_{\mu\nu}$, we first calculate the curvature tensor $\Omega_{\mu'\nu'}$ and then perform a basis transformation. This latter method is mathematically equivalent to the former, but is simpler to perform in a technical sense. Using (\ref{old_connection}) and the definition (\ref{cur_ten}), we obtain:
\begin{equation}
\Omega_{i'j'}=\frac{1}{2}\epsilon_{i'j'k'}\frac{\rho^{k'}}{|\pmb{\rho}|^3},
\label{old_curvature}
\end{equation}
while the other tensor components vanish. By using the tensor transformation identity, as well as the fact that $\Omega_{0'\nu'}$ and $\Omega_{\mu'0'}$ vanish, we get the curvature components necessary for identifying the fields:
\begin{equation}
\Omega_{ij}=\frac{\partial \rho^{i'}}{\partial x^{i}}\frac{\partial \rho^{j'}}{\partial x^{j}}\Omega_{i'j'},
\quad\quad
\Omega_{i0}=-\frac{1}{c^2}\frac{\partial \rho^{i'}}{\partial x^{i}}\frac{\partial \rho^{j'}}{\partial t}\Omega_{i'j'}.
\label{curvature_new}
\end{equation}
Here, $x^i$ denote spatial coordinates $\{x,y,z\}$, while $\rho^{i'}$ denote coordinates $\{\rho^1,\rho^2,\rho^3\}$. In the case of general 1D motion, this transformation further simplifies due to the trivial relations $\rho^1=x$ and $\rho^2=y$. However, unlike the special constant velocity case in Appendix \ref{constantv_appendix}, the coordinate $\rho^3=\gamma(t_r) [z-w_z(t_r)-v_z(t_r)(t-t_r)]$ is a function of both time and all three spatial coordinates due to the presence of the retarded time. Using (\ref{curvature_new}), and identifying the fields via (\ref{EB_Omega}), we obtain the field components:
\begin{align}
E_x &= \frac{\partial\rho^3}{\partial z}\Omega_{2'3'}, &
B_x &= -\frac{1}{c^2}\frac{\partial\rho^3}{\partial t}\Omega_{1'3'}, \nonumber\\
E_y &= \frac{\partial\rho^3}{\partial z}\Omega_{3'1'}, &
B_y &= -\frac{1}{c^2}\frac{\partial\rho^3}{\partial t}\Omega_{2'3'}, 
\label{EB1D_not_simplified}\\
E_z &= \Omega_{1'2'}-\frac{\partial\rho^3}{\partial y}\Omega_{3'1'}-
\frac{\partial\rho^3}{\partial x}\Omega_{2'3'}, &
B_z &= 0. \nonumber
\end{align}
The coordinate $\rho^3$ depends explicitly on $z$ and $t$, and has an additional dependence on $\{t,x,y,z\}$ through $t_r$. Using the chain differentiation rule and identity (\ref{old_curvature}), the expressions for the fields become:
\begin{align}
E_x &= \frac{1}{2}\frac{x}{|\pmb{\rho}|^3}\left[\gamma+\frac{\partial t_r}{\partial z}\frac{\partial \rho^3}{\partial t_r}\right], &
B_x &= \frac{1}{2c^2}\frac{y}{|\pmb{\rho}|^3}\left[-\gamma 
v_z + \frac{\partial t_r}{\partial t}\frac{\partial \rho^3}{\partial t_r}\right], \nonumber\\
E_y &= \frac{1}{2}\frac{y}{|\pmb{\rho}|^3}\left[\gamma+\frac{\partial t_r}{\partial z}\frac{\partial \rho^3}{\partial t_r}\right], &
B_y &= \frac{1}{2c^2}\frac{-x}{|\pmb{\rho}|^3}\left[-\gamma 
v_z + \frac{\partial t_r}{\partial t}\frac{\partial \rho^3}{\partial t_r}\right], 
\label{EB1D_simplified}\\
E_z &= \frac{1}{2}\frac{1}{|\pmb{\rho}|^3}\left[\rho^3-\left(y \frac{\partial t_r}{\partial y}+x \frac{\partial t_r}{\partial x}\right)\frac{\partial \rho^3}{\partial t_r}\right], &
B_z &= 0. \nonumber
\end{align}
The required derivatives are given below. Derivatives of retarded time were obtained by implicitly differentiating 
its defining relation $c(t-t_r)=|{\bf r}-{\bf w}(t_r)|$:
\begin{equation}
\frac{\partial t_r}{\partial t} = \frac{s}{s-\pmb{\beta}\cdot\pmb{s}},
\quad
\frac{\partial t_r}{\partial x^i} = -\frac{1}{c}\frac{s^i}{s-\pmb{\beta}\cdot\pmb{s}},
\quad
\frac{\partial \rho^3}{\partial t_r} = -\frac{1}{c}\gamma^3 a_z\left[s-\pmb{\beta}\cdot\pmb{s}\right],
\label{derivatives_1D}
\end{equation}
where vector $\pmb{s}$ denotes the spatial part of $s^{\mu}$, and $s\equiv |\pmb{s}|$, 
and $\pmb{\beta}={\bf v}/c$. 
In order to eliminate $|\pmb{\rho}|$ in the denominator, we use the identity
\begin{equation}
|\pmb{\rho}|=\rho^0=\gamma (s-\pmb{\beta}\cdot\pmb{s}),
\label{rho_causal}
\end{equation}
which is valid for retarded solutions. 
Combining results (\ref{EB1D_simplified}), (\ref{derivatives_1D}) and (\ref{rho_causal}), and condensing the components into a vector notation yields:
\begin{align}
\bf{E} &= \frac{1}{2}\frac{s}{(\pmb{s}\cdot\pmb{u})^3}\left[(c^2-v_z^2) \pmb{u} + 
\pmb{s}\times(c{\pmb{\hat{s}}}\times(a_z{\pmb{\hat{z}}}))\right], 
\label{E1D_appendix}\\
\bf{B} &= \frac{1}{c}\pmb{\hat{s}}\times\bf{E}.
\label{B1D_appendix}
\end{align}
These fields are the same as those resulting from Maxwell equations for 1D motion, up to a multiplicative constant $q/2\pi\epsilon_0$.

\section{Curvilinear charge motion in a plane}
\label{2D_appendix}

In this section we discuss general 2D motion of a charge in the $xy$ plane. 
The calculation is identical to that in Appendix \ref{1D_motion_appendix}, 
with Eqs. (\ref{curvature_new}) being equally valid. 
For simplicity, here we only demonstrate calculation of the $E_z$ field component; the 
other field components can be obtained analogously. 
The Lorentz transformation $\Lambda$ relating displacement 4-vectors 
$\rho^{\mu}$ and $s^{\mu}$, that is, $\rho^\mu=\Lambda^\mu_\nu s^\nu$, is 
now a boost followed by rotation, $\Lambda=RB$, or more specifically:
\begin{equation}
\begin{pmatrix}
\rho^0 \\ \rho^1 \\ \rho^2 \\ \rho^3
\end{pmatrix}
=
\begin{pmatrix}
1 & 0 & 0 & 0 \\
0 & C & S & 0 \\
0 & -S & C & 0 \\
0 & 0 & 0 & 1 
\end{pmatrix}
\begin{pmatrix}
\gamma & -\gamma \beta_x & -\gamma \beta_y & 0 \\
-\gamma \beta_x & \epsilon_x & \delta & 0 \\
-\gamma \beta_y & \delta & \epsilon_y & 0 \\
0 & 0 & 0 & 1 
\end{pmatrix}
\begin{pmatrix}
s^0 \\ s^1 \\ s^2 \\ s^3
\end{pmatrix}.
\label{2D_matrix}
\end{equation}
Here, we have introduced abbreviations: $\epsilon_i \equiv 1+(\gamma-1)\frac{\beta_i^2}{\pmb{\beta}^2}$, $\delta \equiv (\gamma-1)\frac{\beta_x\beta_y}{\pmb{\beta}^2}$, $S \equiv \sin{\theta_z}$ and $C \equiv \cos{\theta_z}$. 
The rotation angle is $\theta_z(t_r)=\int^{t_r} \omega_{Th}(t') dt'$, where the Thomas precession frequency 
is orthogonal to the $xy$ plane: 
${\pmb \omega}_{Th}(t') = \omega_{Th}(t') {\bf \hat z}$. 
At the end of the calculation one finds that the resulting fields do not depend explicitly on $\theta_z$, 
but rather only on its derivative $\frac{d\theta_z}{dt_r} = \omega_{Th}(t_r)= 
\frac{1}{c^2}\frac{\gamma^2}{\gamma+1}(a_x v_y - v_x a_y)$. 
This property is consistent with Maxwell's theory, that is, the fields at $x^\mu$ may only depend on the velocity and acceleration 
of the charge evaluated in a single point in spacetime - the retarded position of the charge. 
Any explicit field dependency on $\theta_z$ would violate this property, as $\theta_z$ contains the "history" of motion.

For later comparison, here we write the component $E_z$ obtained from Maxwell equations for 2D charge motion in $xy$ plane:
\begin{equation}
E_z = \frac{1}{2}\frac{z}{\gamma^3 \alpha^3}\left[\gamma + \frac{1}{c^2}\gamma^3\left(a_x s^1 + a_y s^2\right)\right]
\label{2D_maxwell}
\end{equation}
where $s=s^0=|\pmb{s}|$, and we have introduced the abbreviation $\alpha \equiv s-\pmb{s}\cdot\pmb{\beta}$. 
It is convenient to rewrite the second term inside the square brackets by multiplying it with $\alpha / \alpha=1$:
\begin{multline}
E_z = \frac{1}{2}\frac{z}{\gamma^3 \alpha^3}\biggl[\gamma + \frac{1}{c^2 \alpha}\gamma^3\biggl(a_x s^1s^0 + a_y s^2s^0 - a_x\beta_x (s^1)^2 - a_y\beta_y (s^2)^2 \\ - (a_x\beta_y + a_y\beta_x)s^1s^2 \biggr) \biggr] \equiv \frac{1}{2} \frac{z}{\gamma^3\alpha^3} (\star\star)
\label{2D_maxwell_expanded}
\end{multline}
The symbol $(\star\star)$ denotes the contents of the square brackets.

Now we return to our theory. The components of vector $\pmb{\rho}$ are:
\begin{align}
\rho^1 &= -\gamma (C\beta_x+S\beta_y)s^0 + (C\epsilon_x+S\delta)s^1 + (C\delta+S\epsilon_y)s^2 \label{2d_rho1}\\
&\equiv \rho^{1(0)}s^0 + \rho^{1(1)}s^1 + \rho^{1(2)}s^2, \nonumber \\
\rho^2 &= -\gamma (-S\beta_x+C\beta_y)s^0 + (-S\epsilon_x+C\delta)s^1 + (-S\delta+C\epsilon_y)s^2 \label{2d_rho2}\\
&\equiv \rho^{2(0)}s^0 + \rho^{2(1)}s^1 + \rho^{2(2)}s^2, \nonumber \\
\rho^3 &= s^3 = z \label{2d_rho3}.
\end{align}
Here, we have introduced the coefficients $\rho^{i(j)}$ as abbreviations for the expressions multiplying $s^0$, $s^1$ and $s^2$: 
$\rho^{1(0)}=-\gamma (C\beta_x+S\beta_y)$, $\rho^{1(1)}=C\epsilon_x+S\delta$, etc. 
In order to calculate $E_z$, we use equation (\ref{curvature_new}) for $\Omega_{12}$. In this case, only two terms survive:
\begin{equation}
E_z = \left(\frac{\partial \rho^1}{\partial x}\frac{\partial \rho^2}{\partial y} -\frac{\partial \rho^2}{\partial x}\frac{\partial \rho^1}{\partial y}\right)\Omega_{1'2'} = \frac{1}{2} \frac{z}{\gamma^3\alpha^3} \left(\frac{\partial \rho^1}{\partial x}\frac{\partial \rho^2}{\partial y} -\frac{\partial \rho^2}{\partial x}\frac{\partial \rho^1}{\partial y}\right)
\equiv \frac{1}{2} \frac{z}{\gamma^3\alpha^3} (\star);
\label{2D_Ez_original}
\end{equation}
$\rho^1$ and $\rho^2$ depend explicitly on $x$ and $y$ through $s^1$ and $s^2$, while also having an additional dependence due to the presence of $t_r$. Using the differentiation chain rule, we get:
\begin{align}
(\star) &= \left[\rho^{1(1)}+\frac{\partial t_r}{\partial x}\dot{\rho^1}\right]
\left[\rho^{2(2)}+\frac{\partial t_r}{\partial y}\dot{\rho^2}\right] -
\left[\rho^{1(2)}+\frac{\partial t_r}{\partial y}\dot{\rho^1}\right]
\left[\rho^{2(1)}+\frac{\partial t_r}{\partial x}\dot{\rho^2}\right] 
\label{2D_star} \\
&= \left[\rho^{1(1)}\rho^{2(2)}-\rho^{1(2)}\rho^{2(1)}\right]+
\dot{\rho^1}\left[\frac{\partial t_r}{\partial x}\rho^{2(2)}-\frac{\partial t_r}{\partial y}\rho^{2(1)}\right] +
\dot{\rho^2}\left[\frac{\partial t_r}{\partial y}\rho^{1(1)}-\frac{\partial t_r}{\partial x}\rho^{1(2)}\right], \label{2D_star_simplified}
\end{align}
where a dot above a symbol denotes a partial derivative with respect to retarded time. The first square bracket in (\ref{2D_star_simplified}) equals $\gamma$. The derivatives of $\rho^1$ and $\rho^2$ with respect to $t_r$ are:
\begin{equation}
\dot{\rho^{j}} = \dot{\rho^{j(0)}}s^0 + \dot{\rho^{j(1)}}s^1 + \dot{\rho^{j(2)}}s^2 - c[\rho^{j(0)} + \beta_x \rho^{j(1)} + \beta_y \rho^{j(2)}], \ j\in\{1,2\}. 
\label{2d_rho_derivative}
\end{equation}
Derivatives of $t_r$ with respect to $x$ and $y$ are given in (\ref{derivatives_1D}). After inserting (\ref{2d_rho_derivative}) and (\ref{derivatives_1D}) into (\ref{2D_star_simplified}), the terms inside square brackets in (\ref{2d_rho_derivative}) exactly cancel out. 
By grouping the surviving terms, we obtain:
\begin{align}
(\star) = \gamma + \frac{1}{c\alpha}\biggl(
&[\dot{\rho^{2(0)}}\rho^{1(2)}-\dot{\rho^{1(0)}}\rho^{2(2)}] s^1s^0 + 
[\dot{\rho^{1(0)}}\rho^{2(1)}-\dot{\rho^{2(0)}}\rho^{1(1)}] s^2s^0 + \label{2D_star_final} \\
&[\dot{\rho^{2(1)}}\rho^{1(2)}-\dot{\rho^{1(1)}}\rho^{2(2)}] (s^1)^2 + 
[\dot{\rho^{1(2)}}\rho^{2(1)}-\dot{\rho^{2(2)}}\rho^{1(1)}] (s^2)^2 + \nonumber\\
&[\dot{\rho^{1(1)}}\rho^{2(1)}-\dot{\rho^{1(2)}}\rho^{2(2)}
+\dot{\rho^{2(2)}}\rho^{1(2)}-\dot{\rho^{2(1)}}\rho^{1(1)}] s^1s^2 \biggr). \nonumber
\end{align}
Comparing (\ref{2D_maxwell_expanded}) and (\ref{2D_Ez_original}), our goal is to prove that $(\star)$ equals $(\star\star)$. Within their definitions, $s^0$, $s^1$ and $s^2$ are the only quantities that contain an explicit dependence on $x$ or $y$. Therefore, for $(\star)$ and $(\star\star)$ to be equal for an arbitrarily chosen point in spacetime, the coefficients $C_{\mu\nu}$ multiplying the products $s^{\mu} s^{\nu}$ must also be the same in both expressions. Note that the coefficient index notation is unrelated to tensor notation, it just serves to signify which product the coefficient belongs to. We have effectively split the problem into 5 smaller problems, all of which can be solved in an analogous way. As a quick demonstration, we consider the coefficient multiplying $s^1s^0$, which is equal to:
\begin{align}
C_{10} &=\frac{1}{c\alpha}[\dot{\rho^{2(0)}}\rho^{1(2)} -\dot{\rho^{1(0)}}\rho^{2(2)}] \label{2D_C10_original} \\
&= \frac{1}{c\alpha}\left[(C\delta+S\epsilon_y)\frac{d}{dt_r}(\gamma(S\beta_x-C\beta_y))+(-S\delta+C\epsilon_y)\frac{d}{dt_r}(\gamma(C\beta_x+S\beta_y))\right] \label{2D_C10_plugged_in} \\
&= \frac{1}{c\alpha}\left[\gamma^2\beta_y(\dot{S}C-S\dot{C})+\left(\epsilon_y\frac{d}{dt_r}(\gamma\beta_x)-\delta\frac{d}{dt_r}(\gamma \beta_y)\right)\right] \label{2D_C10_simplified}
\end{align}
As expected, the remaining quantity $\dot{S}C-S\dot{C}$ is equal to $\frac{d\theta_z}{dt_r}$, while all terms containing explicit dependence on $\theta_z$ exactly cancel out. Thus, we see that the fields depend on the derivative of the angle $\frac{d\theta_z}{dt_r}$.
The remaining functions in (\ref{2D_C10_simplified}) depend on $t_r$ through $\beta_x$ and $\beta_y$. Evaluating the derivatives via the chain rule, and plugging in the definition of $\frac{d\theta_z}{dt_r}$, we get the final result:
\begin{equation}
C_{10} = \frac{\gamma^3}{c^2\alpha}a_x,
\label{2D_C10_final}
\end{equation}
which is exactly equal to the coefficient in $(\star\star)$. An analogous procedure can be completed for the remaining 4 coefficients to conclude that the expressions for $E_z$ are, indeed, equal.


\begin{thebibliography}{99} %%%%%%%%%%%%%%%%%%%%%%%%%%%%%%%%%%%%%%%%%%%%%%%%%%%%%%%%%%%%%%%%%%%%%%%%%%%%%%%


\bibitem{Dirac1931}
P. A. M. Dirac, 
{\em Quantised singularities in the electromagnetic field},
Proc. R. Soc. London {\bf A133}, 60 (1931).

\bibitem{Dirac1948}
P. A. M. Dirac, 
{\em The theory of magnetic poles},
Phys. Rev {\bf 74}, 60 (1948).

\bibitem{Schwinger1969}
J. Schwinger, 
{\em A magnetic model of matter},
Science {\bf 165}, 757 (1969).

\bibitem{Witten1979}
E. Witten, 
{\em Dyons of Charge $e\theta/2\pi$},
Phys. Lett. B {\bf 86}, 283 (1979).

\bibitem{Wilczek1982}
F. Wilczek, 
{\em Remarks on Dyons},
Phys. Rev. Lett. {\bf 48}, 1146 (1982).

\bibitem{tHooft1974}
G. 't Hooft,
{\em Magnetic monopoles in unified gauge theories},
Nuclear Physics B {\bf 79}, 276 (1974).

\bibitem{Polyakov1974}
A.M. Polyakov, 
{\em Particle Spectrum in the Quantum Field Theory}, 
JETP Letters {\bf 20} 194 (1974).

\bibitem{Goldhaber1990}
A.S. Goldhaber and W.P. Trower,
{\em Resource Letter MM-1: Magnetic monopoles},
American Journal of Physics {\bf 58}, 429 (1990).

\bibitem{Chern1944}
Shiing-Shen Chern,  
{\em A Simple Intrinsic Proof of the Gauss-Bonnet Formula for Closed Riemannian Manifolds}, 
The Annals of Mathematics {\bf 45}, 747 (1944). 

\bibitem{Griffiths2013}
D.J. Griffiths, 
{\em Introduction to Electrodynamics}
(Pearson, London, 2013).

\bibitem{Jackson1999}
J.D. Jackson, 
{\em Classical Electrodynamics}
(John Wiley and Sons, New York, 1999)

\bibitem{Dirac1938}
P. A. M. Dirac, 
{\em Classical theory of radiating electrons},
Proc. R. Soc. London {\bf A167}, 148 (1938).

\bibitem{Wheeler1945}
J. A. Wheeler and R. P. Feynman, 
{\em Interaction with the absorber as the mechanism of radiation},
Rev. Mod. Phys. {\bf 17}, 157 (1945).

\bibitem{Wheeler1949}
J. A. Wheeler and R. P. Feynman, 
{\em Classical electrodynamics in terms of direct interparticle action},
Rev. Mod. Phys. {\bf 21}, 425 (1949).

\bibitem{Beil1975}
R. G. Beil, 
{\em Alternate formulations of classical electrodynamics},
Phys. Rev.D {\bf 12}, 2266 (1975).

\bibitem{Feynman1966}
R. P. Feynman, 
{\em The development of the space‐time view of quantum electrodynamics}, 
Physics Today {\bf 19}, 31 (1966).

\bibitem{Weyl1929}
H. Weyl, 
{\em Elektron und Gravitation. I},
Z. Phys. {\bf 56}, 330 (1929).

\bibitem{Berry1984}
M. V. Berry, 
{\em Quantal Phase Factors Accompanying Adiabatic Changes},
Proc. R. Soc. London A {\bf 392}, 45 (1984).

\bibitem{Thomas1927}
L.H. Thomas B.A.,
{\em The kinematics of an electron with an axis}, 
The London, Edinburgh, and Dublin Philosophical Magazine and Journal of Science {\bf 3}, 1 (1927). 

\bibitem{Wu1975}
T.T. Wu and C.N. Yang, 
{\em Concept of nonintegrable phase factors and global formulation of gauge fields},
Phys. Rev.D {\bf 12}, 3845 (1975).

\bibitem{Turner2013}
A. M. Turner and A. Vishwanath, 
{\em Beyond Band Insulators: Topology of Semi-metals and Interacting Phases}, 
arXiv:1301.0330.

\bibitem{Lu2015}
L. Lu, Z. Wang, D. Ye, L. Ran, L. Fu, J.D. Joannopoulos and M. Solja\v{c}i\'{c}, 
{\em Experimental observation of Weyl points},
Science {\bf 349}, 622 (2015). 

\bibitem{Xu2015}
Xu S-Y {\em et al.},
{\em Discovery of a Weyl fermion semimetal and topological Fermi arcs},
Science {\bf 349} 613 (2015). 

\bibitem{Lv2015}
B.Q. Lv {\em et al.}, 
{\em Discovery of Weyl semimetal TaAs}, 
Phys. Rev. X {\bf 5}, 031013 (2015).

\bibitem{Dubcek2015}
T. Dub\v{c}ek, C.J. Kennedy, L. Lu, W. Ketterle, M. Solja\v{c}i\'{c}, and H. Buljan, 
{\em Weyl Points in Three-Dimensional Optical Lattices: Synthetic Magnetic Monopoles in Momentum Space}, 
Phys. Rev. Lett. {\bf 114}, 225301 (2015).


\bibitem{Morgan1995}
P. Morgan,
{\em The Massless Dirac Equation, Maxwell’s Equation, and the Application of Clifford Algebras},
in R. Ablamowitz and P. Lounesto (eds), {\em Clifford Algebras and Spinor Structures}
(Springer Science, Dordrecht, 1995). 

\bibitem{Rozzi2009}
T. Rozzi, D. Mencarelli, and L. Pierantoni,
{\em Deriving Electromagnetic Fields From the Spinor Solution of the Massless Dirac Equation},
IEEE Trans. Microw. Theory Tech. {\bf 57} 2907 (2009).

\bibitem{Bialynicki1987}
I. Bialynicki-Birula and Z. Bialynicka-Birula, 
{\em Berry’s phase in the relativistic theory of spinning particles},
Phys. Rev. D {\bf 35}, 2383 (1987).

\bibitem{VanMechelen2019}
T. Van Mechelen and J. Zubin
{\em Photonic Dirac monopoles and skyrmions: spin-1 quantization},
Optical Materials Express {\bf 9}, 95 (2019).


\end{thebibliography}
\end{document}